\documentstyle[preprint,aps]{revtex}

\begin{document}
\draft
\tighten

\preprint{\vbox{\hbox{IMPERIAL/TP/94-95/18}
\hbox{CWRU-P2-1995}
\hbox{NI94043}
\hbox{hep-ph/9503223}}}

\title{Defect Production in Slow First Order Phase Transitions}

\author{Julian Borrill and T.W.B. Kibble}
\address{Blackett Laboratory, Imperial College, London SW7 2BZ,
United Kingdom and\\
Isaac Newton Institute for Mathematical Sciences, Cambridge CB3 0EH,
United Kingdom}
\author{Tanmay Vachaspati}
\address{Department of Physics, Case Western Reserve University\\
10900 Euclid Ave., Cleveland OH 44106-7079, and\\
Isaac Newton Institute for Mathematical Sciences, Cambridge CB3 0EH,
United Kingdom}
\author{Alexander Vilenkin}
\address{Institute of Cosmology,
Department of Physics and Astronomy, \\
Tufts University, Medford MA 02155, USA and\\
Isaac Newton Institute for Mathematical Sciences, Cambridge CB3 0EH,
United Kingdom}

\date{\today}

\maketitle

\begin{abstract}
We study the formation of vortices in a U(1) gauge theory
following a first-order transition proceeding by bubble nucleation, in
particular the effect of a low velocity of expansion of the bubble
walls.   To do this, we use a two-dimensional model in which bubbles
are nucleated at random points in a plane and at random times and then
expand at some velocity $v_{\rm b}<c$.  Within each bubble, the phase
angle is assigned one of three discrete values.  When bubbles collide,
magnetic `fluxons' appear: if the phases are different, a
fluxon--anti-fluxon pair is formed.  These fluxons are eventually
trapped in three-bubble collisions when they may annihilate or form
quantized vortices.  We study in particular the effect of changing the
bubble expansion speed on the vortex density and the extent of
vortex--anti-vortex correlation.
  \end{abstract}


\narrowtext

\section{Introduction}

Cosmic strings are topological defects which may have been formed at a
high-temperature phase transition very early in the history of the
Universe \cite{reviews}.  To estimate their observational
consequences, we need to follow the evolution of a network of cosmic
strings, either analytically or numerically.  In either case the
starting point must be some estimate of the initial string density
shortly after the symmetry-breaking phase transition.  There has
recently been considerable debate about this question, particularly in
the case of gauge theories \cite{RudSri93,HinDavBra94}.

To be specific, we consider a spontaneously broken Abelian U(1) gauge
theory, with a complex scalar field $\phi$.  At zero temperature,
there is a degenerate ground state in which $\phi$ has a vacuum
expectation value of fixed magnitude, but arbitrary phase.  The
symmetry is restored above some critical temperature $T_{\rm c}$.  The
conventional picture of defect formation is this \cite{Kib76}: as the
Universe cools, the phases in sufficiently far-separated regions are
uncorrelated; if the phase change around a large loop in space is a
non-zero multiple of $2\pi$, then a string or strings must pass
through it.  If the transition is second-order, we may model the
effect by considering the space as made up (at some temperature
slightly below $T_{\rm c}$) of separate domains whose size is
determined by the length scales of the microphysics \cite{KibVil95a},
and supposing that within each domain the phase is chosen randomly and
independently.  Across the boundary between two neighboring domains,
the phase is assumed to interpolate smoothly between the two values,
along the shortest path (this is called the `geodesic rule')
\cite{VacVil84}.  On the line where three domain meet, a string is
trapped if the net phase change around the line is $\pm2\pi$.

Here, we shall be particularly concerned with the case where the
transition is {\it first-order}, proceeding by bubble nucleation.  In
this case, the expanding Universe supercools, remaining in the
symmetric phase below the critical temperature.  When it has cooled
sufficiently, bubbles of the true-vacuum phase begin to nucleate and
expand until they eventually percolate and fill the whole of space.
We may then reasonably assume that the phase within each bubble is
chosen randomly and independently.

There is of course a complication here, emphasized in a recent paper
by Rudaz and Srivastava \cite{RudSri93}:  because of the gauge
invariance, the phase of the field is not well-defined.  Indeed unless
we have fixed the gauge, even the {\it relative\/} phase between
different bubbles is not meaningful.  Is it then correct, in order
to estimate the initial string density, to use a model in which a
phase is randomly assigned in each bubble and strings are trapped
when three bubbles meet if the net phase change is $\pm2\pi$?

In a previous paper \cite{KibVil95b}, two of us have analyzed the
meaning of relative phase and the process of phase equilibration in a
bubble collision.  The relative phase can be given a gauge-invariant
definition, in terms of the line integral of the covariant
derivative.  In general, that definition is path-dependent, but the
{\it initial\/} phase difference is unambiguous, provided we may
assume that the electromagnetic field is initially zero.  We showed
that, in the usually considered case where the bubble walls expand
almost at the speed of light, phase equilibration always proceeds more
slowly than bubble expansion, so that the simple model just described
is indeed correct.  However, this leaves open the question of what
happens when the bubble walls expand more slowly, due to the damping
effect of the ambient plasma.  This is the
question we aim to address in the present paper.

When two bubbles with different phases meet, a current flows across
the interface.  The phase difference between the two bubbles (along
their line of centres) oscillates with frequency determined by the
gauge-field mass and decays due to damping by the plasma
\cite{KibVil95b}.  The current in turn generates a loop of magnetic
flux surrounding the collision region.  When the bubbles expand at
close to the speed of light, the radius of the collision region, and
hence of the flux loop, expands faster than $c$.  (In that case, it
is best to think of the flux not as expanding but as being created
and decaying at successively larger radii.)  Where three bubbles
meet, the fluxes in the three separate loops combine.  Of course
the total flux trapped is then either zero or $\pm$a flux quantum.
Because the flux is always tied to the collision region, we can find
out whether a string is trapped merely by examining the initial
phases of the three bubbles.

The situation is very different, however, if the bubble walls
expand more slowly.  Then the rate of expansion of the flux loop may
be {\it less\/} than $c$, in which case it is quite possible that
it might propagate away from the collision region and become
separated from the bubbles that gave it birth.  If that happens, it
would be likely to suppress the number of strings produced and to
affect their distribution --- in particular the relative proportions
of long strings and loops.  One might think that if strings become
rare, then most of them would be in the form of small closed loops
\cite{Vach91}.  If so, that would have a dramatic impact on cosmology.

The actual speed of magnetic flux spreading depends on the plasma
conductivity in the region between the bubbles.  A realistic
simulation of all the processes involved would be very complicated,
and we shall not attempt it here.  Instead, we shall study a simple
two-dimensional model and represent the flux spreading by the
propagation of particle-like `fluxons'.  Although it is obviously
unrealistic in details, we believe that this model includes the
essential features necessary to answer (qualitatively) our central
question: what is the effect of flux spreading on the defect
statistics?

We shall
consider a two-dimensional space, in which circular bubbles nucleate
at random and expand with some velocity $v_{\rm b}$, and in which
vortices may be trapped at the points where bubbles finally
coalesce.  When two bubbles meet, they generate not a loop of flux
but a fluxon--anti-fluxon pair.  For simplicity, we approximate the
U(1) phase angles by three discrete values $(0, \pm2\pi/3)$.  Then
all the fluxons generated carry the same flux (up to a sign).  In
units of the flux quantum, $2\pi /e$, the possible fluxon charges are
$\pm 1/3$.

We assume that so long as the junction points of the colliding bubble
walls move faster than certain critical speed $v_{\rm f} > v_{\rm
b}$,  the fluxons are fixed to them, but when
the junction speed falls below $v_{\rm f}$, the fluxons are freed and
continue to move independently with speed $v_{\rm f}$, bouncing off
any other bubbles they encounter.

If we wait long enough of course the bubbles will percolate and fill
the whole of space, so all fluxons will eventually become trapped.
The gaps between the bubbles will finally close, trapping vortices
with quantized flux, so the fluxons will either annihilate or be
forced together in threes to form vortices.  But of course the total
flux trapped where three bubbles finally coalesce {\it cannot\/} now
be found just by looking at their initial phases.  Some of their
fluxons may have escaped, while others from farther afield may have
wandered in.

When a fluxon passes between two bubbles, it changes the relative
phase between them.  So to follow all the changes in phase
as the system evolves would be a very complicated task.
Fortunately, it is not necessary to do so.  Our strategy will be not
to choose the phases until it is necessary to do so, at bubble
collisions.  When two bubbles collide, there are two distinct cases.
If they belong to {\it disjoint\/} bubble clusters, the relative
phase between these clusters has not yet been fixed, so we make a
random choice, here restricted to the three discrete values.  In
principle it would be possible to trace the evolution of the phase
difference back, following the movements of all the intervening
fluxons, to discover what the initial phase difference was when the
bubbles nucleated, but this is not something we ever need to know.
Whenever it is chosen, the phase difference is random.

The other possibility is that the two bubbles belong to the {\it
same\/} cluster.  Then the collision completes a circuit
within the bubble cluster, usually enclosing a region of the
symmetric phase, or splitting an already enclosed region into two.
(Other cases are described below.)  In that case, the relative phase
between the two colliding bubbles is already in principle fixed by
earlier choices, so we do not have a random choice to make.  In fact,
by consistency, the relative phase must be such as to ensure that the
total flux within the newly enclosed region is an integer number of
flux quanta, i.e., that the net fluxon number is a multiple of
three.  In line with the geodesic rule, we assume the phase
difference is as small as possible consistent with this condition.
In other words, we create at most a single
fluxon--anti-fluxon pair.

The algorithm we adopt is described in the following Section, its
implementation in Section III and the results in Section IV.  We are
particularly interested in examining the dependence of the defect
density  on the velocity ratio $v_{\rm b}/v_{\rm f}$.  If the
bubble-wall velocity is low, one expects the number of defects per
bubble to be reduced.  This is because three-bubble collisions will
less often trap strings, since the phases of the first two bubbles
may have equilibrated before they encounter the third.  In our model
this effect is represented by the escape of the relevant fluxons.  In
the three-dimensional case, another effect could be to change the
ratio of long strings to small loops.  In two dimensions, the
analogue of a small loop is a close vortex--anti-vortex pair, so we
also study the ratio between the mean nearest-neighbor
vortex--anti-vortex distance and the corresponding vortex--vortex
one.  For $v_{\rm b}=v_{\rm f}$, there is strong vortex--anti-vortex
correlation: the ratio is substantially less than one.  But for
$v_{\rm b} < v_{\rm f}$ we shall see that, in contrast to models with
tilted potentials \cite{Vach91}, the reduction in the number of
defects is accompanied by a {\it reduced} vortex--anti-vortex
correlation.  Our conclusions are discussed in Section V.

\section{Algorithm}

In the standard numerical simulations of defect production
\cite{VacVil84} relative phases are assigned at random to sites on a
lattice corresponding to the centres of causally disconnected regions
of true vacuum (either bubbles in a first-order or domains in a
second-order transition). Between these sites the phase is taken to
vary along the shortest path on the vacuum manifold --- the so-called
geodesic rule.  Defects are then formed wherever this geodesic
interpolation between sites generates a topologically nontrivial
path in the vacuum manifold.

For a first-order transition this formalism corresponds to true
vacuum bubbles nucleating simultaneously, equidistant from all their
nearest neighbors. Consequently all collisions between neighboring
bubbles occur simultaneously, and the associated phase differences
are simply given by the differences in the initial assigned phases.
In this paper we refine this approach in two ways. Firstly the
bubbles are nucleated at random times and places, so that their
collisions are also randomly distributed in time and space.
Secondly, we consider the effect of flux spreading on defect
formation, with flux spreading being represented by the propagation
of fluxons.

Our algorithm is therefore as follows:
\begin{enumerate}
\item generate a population of bubble nucleation events distributed
randomly within some finite volume of 2+1 dimensional spacetime.
\item expand these bubbles at some fixed sub-luminal speed $v_{\rm
b}$.
\item at every 2-bubble collision determine the relative phase
difference:--
\begin{enumerate}
\item if the collision does not close off a region of false vacuum,
assign the fluxon pair at the two intersection points at random.
\item if the collision does close off a region of false vacuum,
assign the fluxon pair at the two intersection points so as to round
the total charge within the closed region to the nearest integer.
\end{enumerate}
\item at every intersection point, determine the time at which the
associated fluxon escapes, and follow its subsequent free evolution,
bouncing it off any bubble that it encounters.
\item at every 3-bubble collision, sum the fluxons associated with
the three intersections to give the total defect charge.
\end{enumerate}

All the defects in this approach are ultimately formed at 3-bubble
collisions (or equivalently at the collision of three two-bubble
intersection points). However three bubbles can collide in two ways.
In the first case (Fig.\ 1, called an external collision) all the
collisions occur in the false vacuum and a closed curvilinear
`triangle' of false vacuum is formed whose vertices are the three
two-bubble intersection points. As the bubbles expand this triangle
shrinks to a point, condensing all the charge (both at the
intersection points and in the form of free fluxons trapped in the
closed region) into an integer-charged defect. This is the usual
defect generating mechanism, and the only one available in
lattice-based simulations. Alternatively (Fig.\ 2, called an internal
collision) one of the three collisions occurs within the third bubble.
In this case the three intersection points meet as the internal one
emerges into the false vacuum. Although there is no closed region of
false vacuum, if the two external intersection points carry fluxons of
the same sign, this configuration generates a defect and an
anti-fluxon.  This can be viewed as the production of a virtual
fluxon--anti-fluxon pair as the 1-2 intersection point emerges from
bubble 3.  The fluxon then joins with the 1-3 and 2-3 fluxons to
generate a defect and the anti-fluxon is released.  In the case of
relativistic bubbles, it can be shown that the division of collisions
into internal and external is frame-dependent.  For any internal
collision one can find a frame of reference where it is seen as
external, and vice versa.  Since defect production should be
frame-independent, this suggests that the same rules should be applied
to both types of collisions.

In our algorithm, the defects are held stationary once they are formed
and we have ignored any evolution of the defect gas during the course
of the phase transition. For the specific case we have in mind --- the
breaking of a $U(1)$ gauge symmetry in two dimensions --- there are no
long-range inter-defect forces and the only evolutionary effect would
be due to the initial random velocities of the defects.  Evolution due
to such random velocities would result in some mixing of the defects
and will somewhat decrease the correlations of defect and anti-defect
locations. In the case where there are long range inter-defect forces
or in the case of strings in three dimensions where the string tension
can be the cause of evolution, our results should be used with care.
However, even in these cases, if the mean distance traversed by
defects during the course of the phase transition is not larger than
the inter-defect separation, our results would be applicable.

\section{Implementation}

Our simulation is sufficiently simple that it does not require
discretization of space and time: the bubbles and fluxons can be
evolved completely analytically.  This, however, makes the
implementation quite complicated.    We have first to generate a
population of bubbles, then to calculate when all the key events
(2-bubble collision, fluxon freeing, and 3-bubble collision) occur,
then to time-order these, and only then to work through including them
in the simulation.

\subsection{Preparation}

To generate a population of bubbles we choose some predetermined
number of random events, each lying within the designated simulation
volume. Time-ordering these we reject all those which would
correspond to a bubble being nucleated within another bubble. Finally
we check {\it post hoc} that the remaining bubbles completely fill the
simulation space by the end of the simulation time. We now have a
time-ordered list of bubbles defined by their nucleation events
\begin{equation}
B_{i} \equiv (t_{i}, {\bf x}_{i}).
\end{equation}

Next, we determine the coordinates of the collision event of every
pair of bubbles. Any time-ordered pair $B_{i}, B_{j}$ collide at
$C_{ij} \equiv (t_{ij}, {\bf x}_{ij})$ when
\begin{equation}
 r_{i}(t_{ij}) + r_{j}(t_{ij}) = |\Delta {\bf x}_{ij}|,
\end{equation}
where $r_{i} = v_{\rm b}(t - t_{i})$ is the radius of bubble $B_{i}$
expanding at speed $v_{\rm b}$, and the spatial separation of the bubble
centres is $\Delta {\bf x}_{ij}= {\bf x}_{j}-{\bf x}_{i}$. Solving
for $t_{ij}$ we find
\begin{equation}
t_{ij} = \frac{|\Delta {\bf x}_{ij}|
+ v_{\rm b} (t_{i} + t_{j})} {2 v_{\rm b}}.
\end{equation}
The co-ordinates of the intersection points
associated with collision $C_{ij}$ at any time $t \geq t_{ij}$ are
given by (Fig.\ 3)
\begin{equation}
\label{eIPX}
{\bf x}_{ij}(t) = {\bf x}_{i} + \alpha(t) \hat{\bf n}_{ij}
\pm \beta(t) \hat{\bf n}_{ij}^{\perp},
\end{equation}
where $\hat{\bf n}_{ij}$ is the unit vector along the line of centres
from $B_{i}$ to $B_{j}$, $\hat{\bf n}_{ij}^{\perp}$ is a unit vetor
perpendicular to $\hat{\bf n}_{ij}$, and
\begin{eqnarray}
\alpha(t) = \frac{\Delta {\bf x}_{ij}^{2}+r_{i}^{2}(t)-r_{j}^{2}(t)}
{2 |\Delta {\bf x}_{ij}|}, \nonumber \\
\beta(t) = \sqrt{r_{i}^{2}(t) - \alpha^{2}(t)}.
\end{eqnarray}
Calculating ${\bf x}_{ij}(t_{ij})$ and neglecting any collision
occuring outside the simulation volume, all the remaining 2-bubble
collision events are held in a time-ordered list.

Any fluxon associated with either of the above intersection points
will be released when their speed falls below that of a free fluxon,
$v_{\rm f}$. Differentiating equation (\ref{eIPX}) with respect to
time gives
\begin{equation}
\label{eIPV}
{\bf v}_{ij}(t) = \dot{\alpha}(t) \hat{\bf n}_{ij}
\pm \dot{\beta}(t) \hat{\bf n}_{ij}^{\perp},
\end{equation}
so the intersection points fall below the fluxon speed when
\begin{equation}
|{\bf v}_{ij}(t)| = v_{\rm f}.
\end{equation}
Solving for $t$ we find
\begin{equation}
t = t_{i} + \frac{R_{+}}{v_{\rm b}},
\end{equation}
where $R_{+}$ is the positive root of the equation
\begin{equation}
R^{2} - v_{\rm b} \Delta t_{ij} R
+ \frac{v_{\rm f}^{2}(v_{\rm b}^{2} \Delta t_{ij}^{2}
- \Delta {\bf x}_{ij}^{2})}
{4(v_{\rm f}^{2}-v_{\rm b}^{2})} = 0,
\end{equation}
and the temporal separation of the bubbles is $\Delta t_{ji} =
t_{j}-t_{i}$. The location of the intersection points at this time are
then given by equation (\ref{eIPX}). If such a point lies within
another bubble at this time then any associated fluxon will already
have been involved in a 3-bubble, defect-forming, collision and can be
neglected. Again neglecting any events occuring outside the simulation
volume the remaining potential fluxon-release events are also held in
a time-ordered list.

Ultimately we want the coordinates of all 3-bubble collisions. In
practice we only need consider time-ordered triplets $B_{i}, B_{j},
B_{k}$ for which $C_{ij}, C_{jk}$ and $C_{ki}$ occur within the
simulation time. Such a triplet collides at $D_{ijk} \equiv (t_{ijk},
{\bf x}_{ijk})$ when
\begin{eqnarray}
\label{e3BC}
|{\bf x}_{ijk} - {\bf x}_{i}| & = & r_{i}(t_{ijk}), \nonumber \\
|{\bf x}_{ijk} - {\bf x}_{j}| & = & r_{j}(t_{ijk}), \nonumber \\
|{\bf x}_{ijk} - {\bf x}_{k}| & = & r_{k}(t_{ijk}).
\end{eqnarray}
Solving for $t_{ijk}$ we find
\begin{equation}
a t_{ijk}^{2} + b t_{ijk} + c = 0,
\end{equation}
where
\begin{eqnarray}
a & = & 4 v_{\rm b}^{2} [
v_{\rm b}^{2} (\Delta {\bf x}_{ij} \Delta t_{ik}
- \Delta {\bf x}_{ik} \Delta t_{ij})^2
 - (\Delta {\bf x}_{ij} \times \Delta {\bf x}_{ik})^{2}
], \nonumber \\
b & = & 4 v_{\rm b}^{2}\{
v_{\rm b}^{2} [\Delta {\bf x}_{ij} \cdot \Delta {\bf x}_{ik} \Delta
t_{ij} \Delta t_{ik} (\Delta t_{ij} + \Delta t_{ik}) - \Delta {\bf
x}_{ij}^{2} \Delta t_{ik}^{3}  - \Delta {\bf x}_{ik}^{2} \Delta
t_{ij}^{3}] \nonumber \\
&& + \Delta {\bf x}_{ij}^{2} \Delta {\bf
x}_{ik}^{2} (\Delta t_{ij} + \Delta t_{ik}) - \Delta {\bf x}_{ij}
\cdot \Delta {\bf x}_{ik}  (\Delta {\bf x}_{ij}^{2} \Delta t_{ik} +
\Delta {\bf x}_{ik}^{2} \Delta t_{ij}) \}, \nonumber \\
c & = &
v_{\rm b}^{4} (\Delta {\bf x}_{ij} \Delta t_{ik}^{2} - \Delta {\bf
x}_{ik} \Delta t_{ij}^{2})^2 + \Delta {\bf x}_{ij}^{2} \Delta {\bf
x}_{ik}^{2}  (\Delta {\bf x}_{ij} - \Delta {\bf x}_{ik})^{2}
\nonumber \\
&& + 2 v_{\rm b}^{2} [(\Delta {\bf x}_{ij}^{2} \Delta t_{ik}^{2} +
\Delta {\bf x}_{ik}^{2} \Delta t_{ij}^{2}) \Delta {\bf x}_{ij} \cdot
\Delta {\bf x}_{ik} -  \Delta {\bf x}_{ij}^{2} \Delta {\bf
x}_{ik}^{2}(\Delta t_{ij}^{2} + \Delta t_{ik}^{2})].
\end{eqnarray}
Any solution of this quadratic equation satisfying $t_{ijk} \geq
t_{k}$ then corresponds to a potential defect formation event whose
position is easily established using equation (\ref{eIPX}).  The
correct root is determined by checking for consistency
with the appropriate equation (\ref{e3BC}). Rejecting any events
occuring outside the simulation volume, the remaining 3-bubble
collision events are stored in a time-ordered list.

\subsection{Simulation}

Having generated time-ordered lists of all 2-bubble collisions,
potential fluxon releases, and 3-bubble collisions occuring within the
simulation spacetime volume, we are now in a position to work through
all these events incorporating them into the simulation in the correct
time order. As we step through these events, however, we must also
follow the motion of any free fluxons previously released from slow
intersection points. At the time of its release a fluxon's velocity
is in the direction of motion of the intersection point, and from
equation (\ref{eIPV}) given by
\begin{equation}
{\bf v}_{a} = \frac{v_{\rm f}}{|{\bf v}_{ij}|} {\bf v}_{ij}.
\end{equation}
The free fluxon is then assumed to travel at a constant velocity until
it hits a bubble wall, whereupon it undergoes a relativistic bounce
(Fig.\ 4)
\begin{eqnarray}
\label{eFBB}
v_{1} & = & \frac{2 v_{\rm b} - (1 + v_{\rm b}^{2}) u_{1}}
                 {1 + v_{\rm b}^{2} - 2 u_{1} v_{\rm b}}, \nonumber \\
v_{2} & = & u_{2} \frac{1 - v_{\rm b}^{2}}{1 + v_{\rm b}^{2} - 2 u_{1}
               v_{\rm b}}.
\end{eqnarray}

Before including any event all the free fluxons in the simulation must
be progressively updated to their positions at the time of the event.
For each free fluxon $F_{a} \equiv (t_{a}, {\bf x}_{a}(t_{a}), {\bf
v}_{a}(t_{a}))$ we calculate its collision time $t_{aj}$ with bubble
$B_{j}$ from the condition
\begin{equation}
|{\bf x}_{a}(t_{aj}) - {\bf x}_{j}| = r_{j}(t_{aj}).
\end{equation}
Solving for $t_{aj}$ we find
\begin{equation}
a t_{aj}^{2} + b t_{aj} + c = 0,
\end{equation}
where
\begin{eqnarray}
a & = & {\bf v}_a^{2} - v_{\rm b}^{2}, \nonumber \\
b & = & 2 ({\bf x}_{a}-{\bf x}_{j}) \cdot {\bf v}_{a} - 2t_{a}
{\bf v}_a^{2} + 2t_{j} v_{\rm b}^{2}, \nonumber \\
c & = & ({\bf x}_{a}-{\bf x}_{j})^{2} - 2 t_{a} ({\bf x}_{a}-{\bf
x}_{j}) \cdot {\bf v}_{a} + t_{a}^{2} {\bf v}_a^{2} - t_{j}^{2}
v_{\rm b}^{2},
\end{eqnarray}
and reject as unphysical any roots which are imaginary or in the past
($t_{aj} < t_{a}$ or $t_{aj} < t_{j}$). Taking the smallest physical
collision time the new fluxon position is simply
\begin{equation}
{\bf x}_{a}(t) = {\bf x}_{a}(t_{a}) + (t - t_{a}){\bf v}_{a}(t),
\end{equation}
and the new velocity given by equation (\ref{eFBB}). One further
complication to note is that it is possible for a free fluxon to be
re-captured by a fast intersection point; in this case we simply add
the fluxon charge to whatever is already present at the intersection.
This process is repeated until either the fluxon is re-captured or we
reach the time of the event. Having thus updated every free fluxon we
are now in a position to process the event itself.

If the event is a 2-bubble collision we first determine whether it
closes off a region of false vacuum. This will only occur if the
colliding bubbles are members of the same cluster. Thus if we assign
each bubble a unique cluster number we can immediately tell whether a
collision causes a closure or not. If the collision does not close,
then we assign a fluxon--anti-fluxon pair at random to the two
intersection points and re-number all the members of one cluster with
the cluster number of the other. If the cluster does close, then we
calculate the total charge within the closed region (from both
fluxons trapped at intersection points and free fluxons now trapped
inside the closed false vacuum region) and assign the
fluxon--anti-fluxon pair so as to round the charge in the closed
region to the nearest integer. If the closed region is bounded by
three bubbles then, provided no bubble is nucleated within the region
before it disappears, this integer charge will be that of the
associated defect. Note that because of the possible presence of free
fluxons the defect can have any integer charge, compared with the
$\pm 1$ charges possible in standard simulations.

If the event is a fluxon freeing, then we calculate its initial
position and velocity as above and add it to the free fluxon list,
ready for updating before the next event.

If the event is a 3-bubble collision, then we need to know if it is an
external or an internal collision. If it is an external collision,
then it will have been preceeded by a closing 2-bubble collision and
we will already know the charge associated with the defect. In this
case we simply remove the associated fluxons from the simulation and
add the defect position and charge to the defect list. For internal
collisions free fluxons clearly play no role. We simply sum the
fluxons at the two external intersection points and round the defect
charge to the nearest integer. However in this case we also have to
include the appropriate complementary fluxon at the emerging
intersection point (Fig.\ 2b). If the emerging intersection point is
slow then this fluxon is immediately freed as above.  Furthermore the
convergence of the two external intersection points may reduce a
closed region bounded by four bubbles to one bounded by three, in
which case the new bounded charge should be calculated ready for the
ensuing external 3-bubble collision.

It should be noted that at various points during these
calculation it is necessary to test two real numbers for equality.
Since this can only be done to finite accuracy we have to include a
difference cutoff, below which two numbers are deemed to be equal. To
the extent that this introduces a minimal measureable space and time
interval there is still some discretization inherent in these
simulations.

A few snapshots of the simulation for the case $v_{\rm b} / v_{\rm
f} = 0.5$, with the resulting system of vortices, are shown in Fig.\
5.  The vortex distributions obtained for $v_{\rm b} /v_{\rm f} = 1$
and using the standard lattice simulation of defect formation
\cite{VacVil84} are shown in Fig.\ 6 for comparison.

\section{Results}

We are interested in the variation in the defect statistics with the
ratio $v_{\rm b}/v_{\rm f}$. For each value of this ratio, we must
nucleate a sufficient number of bubbles to fill the simulation space
by the end of the simulation time. However we can then draw the
statistics of interest only from the region sufficiently far away
from the edge of the simulation not to have been affected by the
absence of bubbles beyond the edge. Since the fluxons are taken to
have $v_{\rm f} = 1$, for a 2+1 dimensional simulation of size
$X^{2}$ and duration $T$ this region covers the range $(T, X-T)$ in
each direction. To obtain reasonable statistics we ensure that this
`safe' region contains of the order of 100 bubbles and then
calculate \begin{enumerate}
\item the ratio of the number of defects to the number of bubbles,
$N_{\rm d}/N_{\rm b}$.
\item the ratio of the mean minimum defect--anti-defect separation to
the mean minimum defect--defect separation, $R \equiv \langle D_{\rm
d \bar{d}} \rangle/\langle D_{\rm d d} \rangle$.
\item the fraction of the defects of charge $|Q| = n$ for $n = 1, 2,
\ldots$
\end{enumerate}
The smaller $v_{\rm b}$ is the more bubbles we need to fill the
simulation volume whilst generating 100 `safe' bubbles, and we are
constrained by computing resources to $v_{\rm b} /v_{\rm f} \geq 0.2$. The
results, averaged over 100 runs for each value of $v_{\rm b} /v_{\rm
f}$, are shown in Fig.\ 7.

\section{Discussion}

Our goal in this paper was to study the statistical properties of
the system of vortices formed in a first-order phase transition.  In
particular, we were interested in the dependence of these properties
on the speed of bubble expansion, which is characterized in our model
by the parameter $v_{\rm b} /v_{\rm f}$.  Our main results are
presented in Fig.\ 7 which shows the number of vortices formed per
bubble and the ratio of the average nearest-neighbor
vortex--anti-vortex and vortex--vortex distances, $R = \langle D_{\rm
d \bar{d}} \rangle /\langle D_{\rm dd} \rangle$, as functions of
$v_{\rm b} / v_{\rm f}$.  The ratio $R$ gives a quantitative measure
of the vortex--anti-vortex correlation.

We see, first of all, that the number of vortices decreases as
$v_{\rm b}$ gets smaller (at fixed $v_{\rm f}$).  This is not
difficult to understand.  At low values of $v_{\rm b}$, fluxon escape
prevents the formation of vortices in places where they would
otherwise be formed.  The escaped fluxons are eventually captured,
but they mix with the escaped anti-fluxons, and there is a tendency
for the net flux to cancel.  Annihilation of large groups of fluxons
and anti-fluxons can be seen in Fig.\ 5.

Apart from a decrease in the number of defects, a visual inspection
of vortex distributions in Figs.\ 5e and 6a suggests that the flux
escape decreases correlation between vortices and anti-vortices.  For
$v_{\rm b} = v_{\rm f}$ there is no flux escape, and the distribution in Fig.\
6a
contains many close vortex--anti-vortex pairs, while there are very
few such pairs for $v_{\rm b} = 0.5 v_{\rm f}$.  This trend is
confirmed by the graph in Fig.\ 7 which shows a decrease in
vortex--anti-vortex correlation with a decreasing speed of bubble
walls $v_{\rm b}$.

It is interesting to compare the vortex distribution for $v_{\rm b}
= v_{\rm f}$  with that obtained using a random-phase lattice
simulation (Fig.\ 6b).  The visual appearance of the latter
distribution is quite different, but it also shows a strong
correlation between vortices and anti-vortices.  The nearest
neighbors of almost all vortices are anti-vortices and vice versa.  A
calculation of the ratio $R$ for the lattice simulation gives $R =
0.58$, which is fairly close to the value $R = 0.5$ for $v_{\rm b} =
v_{\rm f}$.  The difference probably arises from the fact that the
lattice imposes a minimum defect--anti-defect separation distance.
It is noteworthy that $R$ decreases with decreasing defect density.
This is in contrast to the `biased' case when the order parameter
potential is tilted \cite{Vach91}. There it is found that $R$
decreases with increasing defect density.

The strong vortex--anti-vortex correlation in lattice simulations has
been known for a long time \cite{Ein80}.  The total magnetic flux
$\Phi$ through a region of size $L$ is proportional to the phase
variation around the region's perimeter.  If the phase varies at
random on the scale of the lattice spacing $\xi$, we have $\Phi
\propto (L/\xi )^{1/2}$.  On the other hand, the number of defects
inside the region is $N \sim (L/\xi)^2$, and an uncorrelated
distribution would give a much larger flux, $\Phi \propto N^{1/2}
\sim L/\xi$.  Essentially the same argument applies to our bubble
simulation, but now the spread in bubble sizes results in a spread in
the nearest-neighbor separations.  This spread is responsible for the
different visual appearance of the two distributions.

The decrease in the vortex--anti-vortex correlation at low bubble
speeds can be easily understood.  Correlations are destroyed when
fluxons escape from the bubble intersections where they originated.
The escaped fluxons form a random gas, and we expect no correlations
on small scales, where fluxons and anti-fluxons had enough time to
randomize.  If $L_{\rm r}$ is the characteristic scale on which
randomization has occured, then we expect magnetic flux fluctuations
to scale as $\Phi \propto N^{1/2}$ for $L < L_{\rm r}$ and as $\Phi
\propto N^{1/4}$ for $L > L_{\rm r}$.

We finally briefly discuss the implications of our results for
defect formation in three-dimensional phase transitions.  As already
mentioned in the Introduction, a close vortex--anti-vortex pair is a
two-dimensional analogue of a small closed loop of string.  Our
results suggest that magnetic flux spreading will decrease the amount
of string in small loops relative to the infinite strings.

If magnetic monopoles are formed in a slow first order phase
transition, we expect a decrease in the monopole density and in the
correlation between monopoles ($M$) and anti-monopoles (${\bar M}$).
For a suitably defined scale $L_{\rm r}$, the magnetic charge
fluctuations will scale as $N^{1/2}$ for $L < L_{\rm r}$ and as
$N^{1/3}$ for $L > L_{\rm r}$.  This randomization of the monopole
distribution can be important in models where monopoles get connected
by strings, particularly in Langacker-Pi-type models \cite{Lang80}
where strings disappear at a subsequent phase transition.  If $M$'s
and ${\bar M}$'s are strongly correlated, as in second-order or fast
first-order transitions, then most of the $M{\bar M}$ pairs get
connected by the shortest possible strings of length $l$ comparable
to the average inter-monopole distance $d$.  Longer strings with $l
\gg d$ are exponentially suppressed \cite{Sik83},  \cite{Cop86}.  For
monopoles formed in a slow first-order transition, the length
distribution of strings can be much broader.  Since the lifetime of
$M{\bar M}$ pairs is determined mainly by the time it takes to
dissipate the energy of the string, the number of monopoles surviving
after the strings disappear can be significantly affected.

\acknowledgments

We are indebted for hospitality to the
Isaac Newton Institute for Mathematical Sciences, Cambridge,
England, where this work was initiated.  The work of AV was supported
in part by the National Science Foundation. TV wishes to thank the
Rosenbaum Foundation for the award of a Fellowship at the Isaac Newton
Institute.

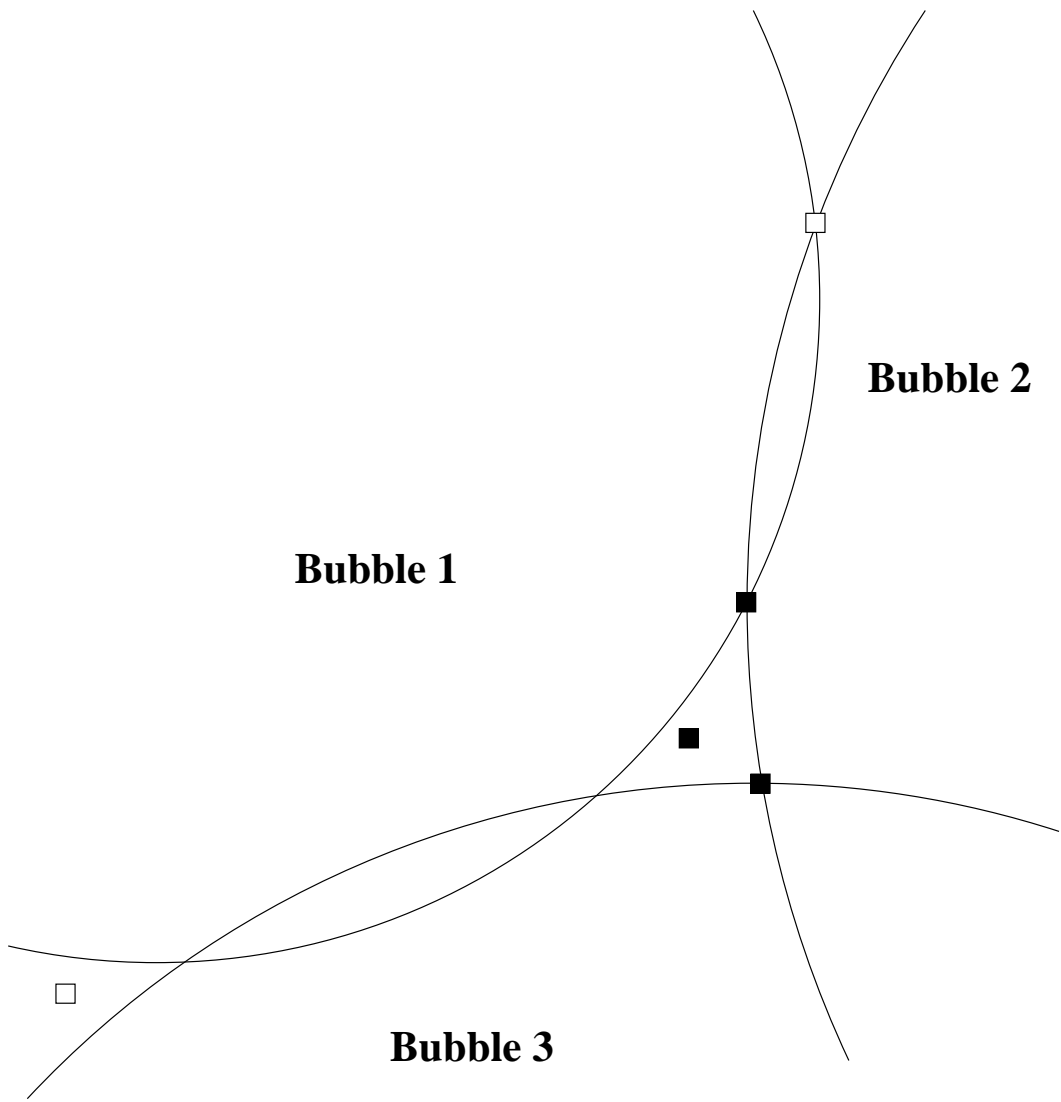
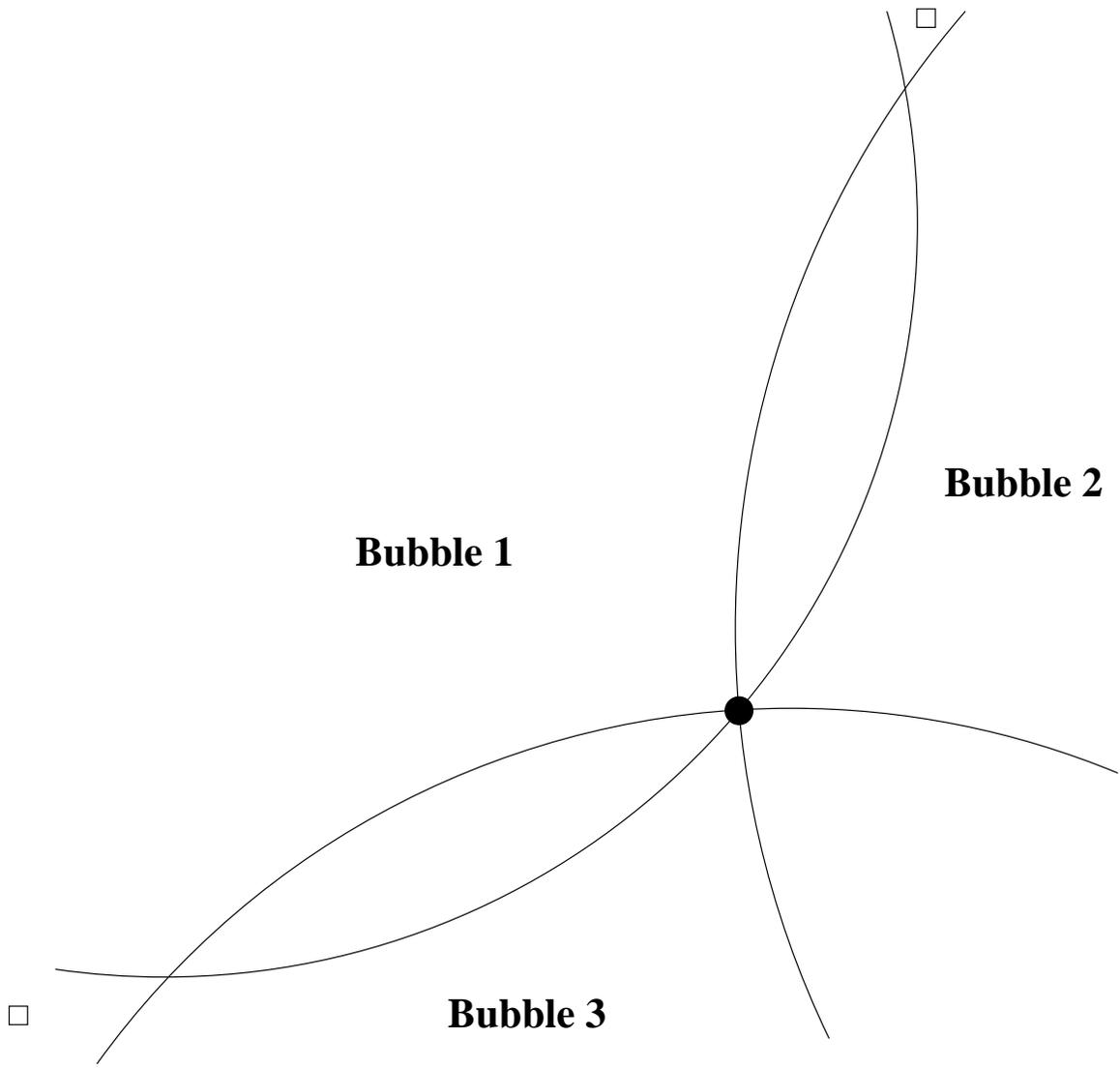
\begin{figure}

\caption{}{An `external' collision of three bubbles (a) before
the collision, (b) at the collision, showing the formation of a
vortex. In all such figures squares represent fluxons and circles
vortices, with black being positively and white negatively charged.}

\end{figure}

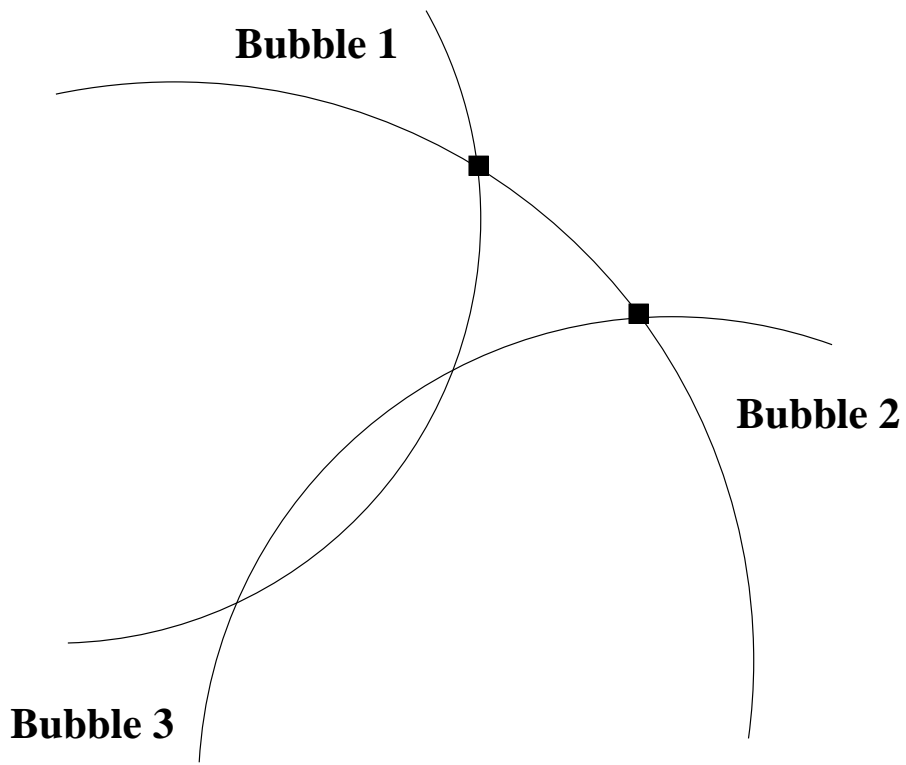
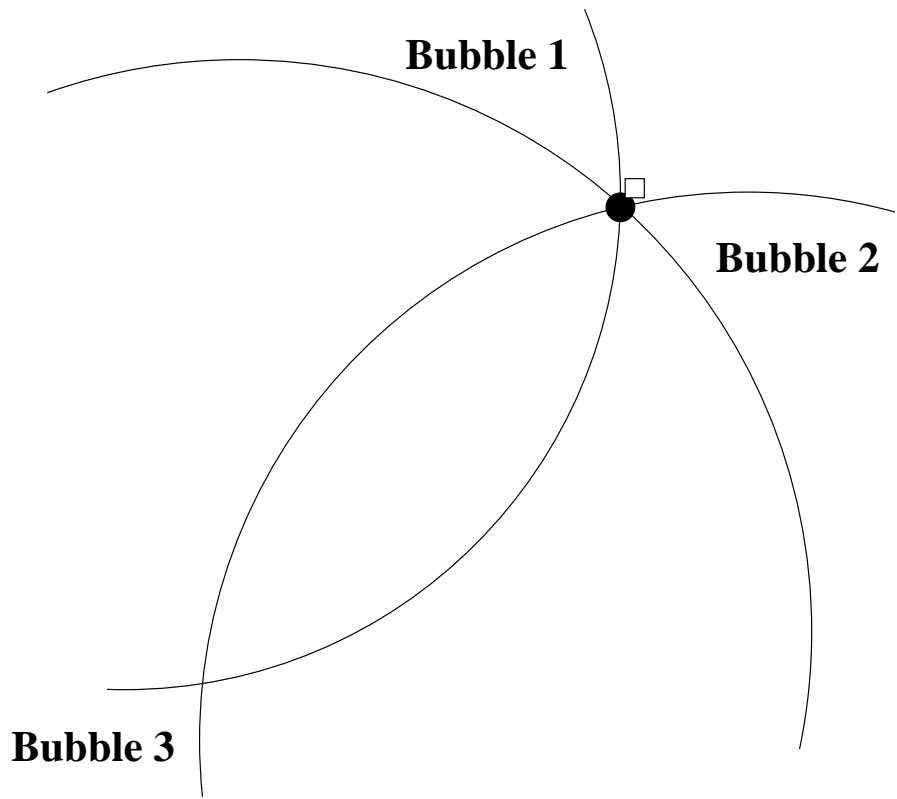
\begin{figure}

\caption{}{An `internal' collision of three bubbles (a) before
the collision, (b) at the collision, showing the formation of a vortex
and a compensating anti-fluxon.}

\end{figure}

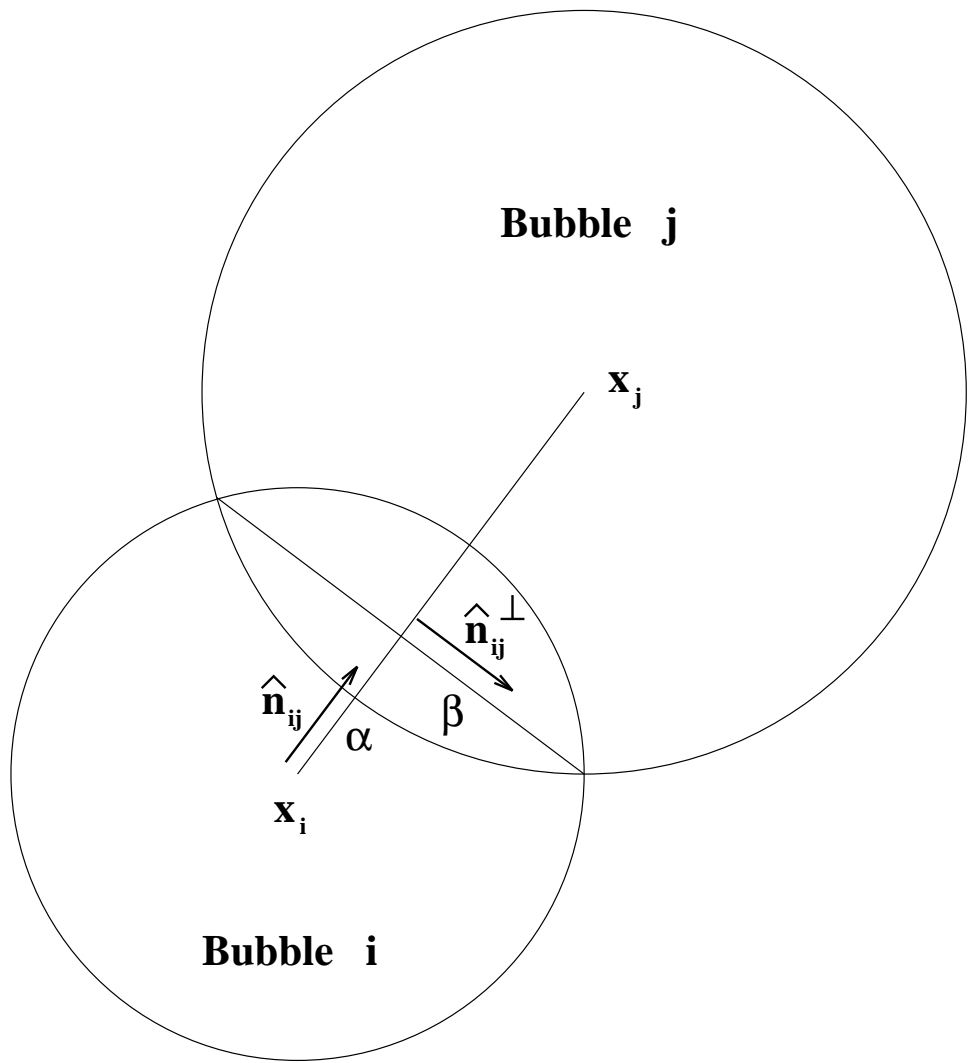
\begin{figure}

\caption{}{The geometry of a two-bubble collision.}

\end{figure}

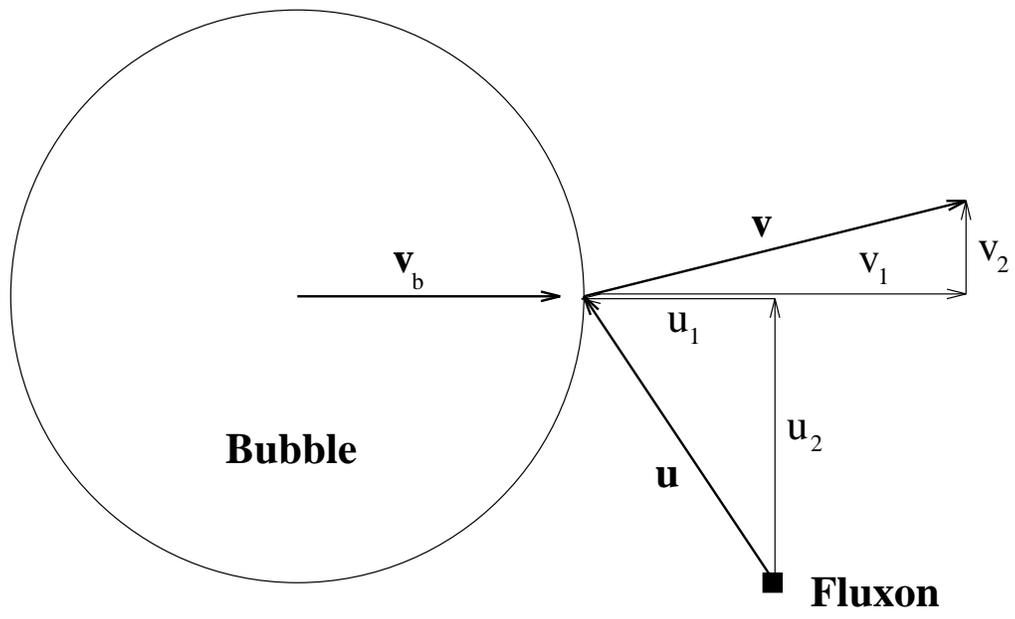
\begin{figure}

\caption{}{The relativistic bounce of a fluxon off a bubble.}

\end{figure}

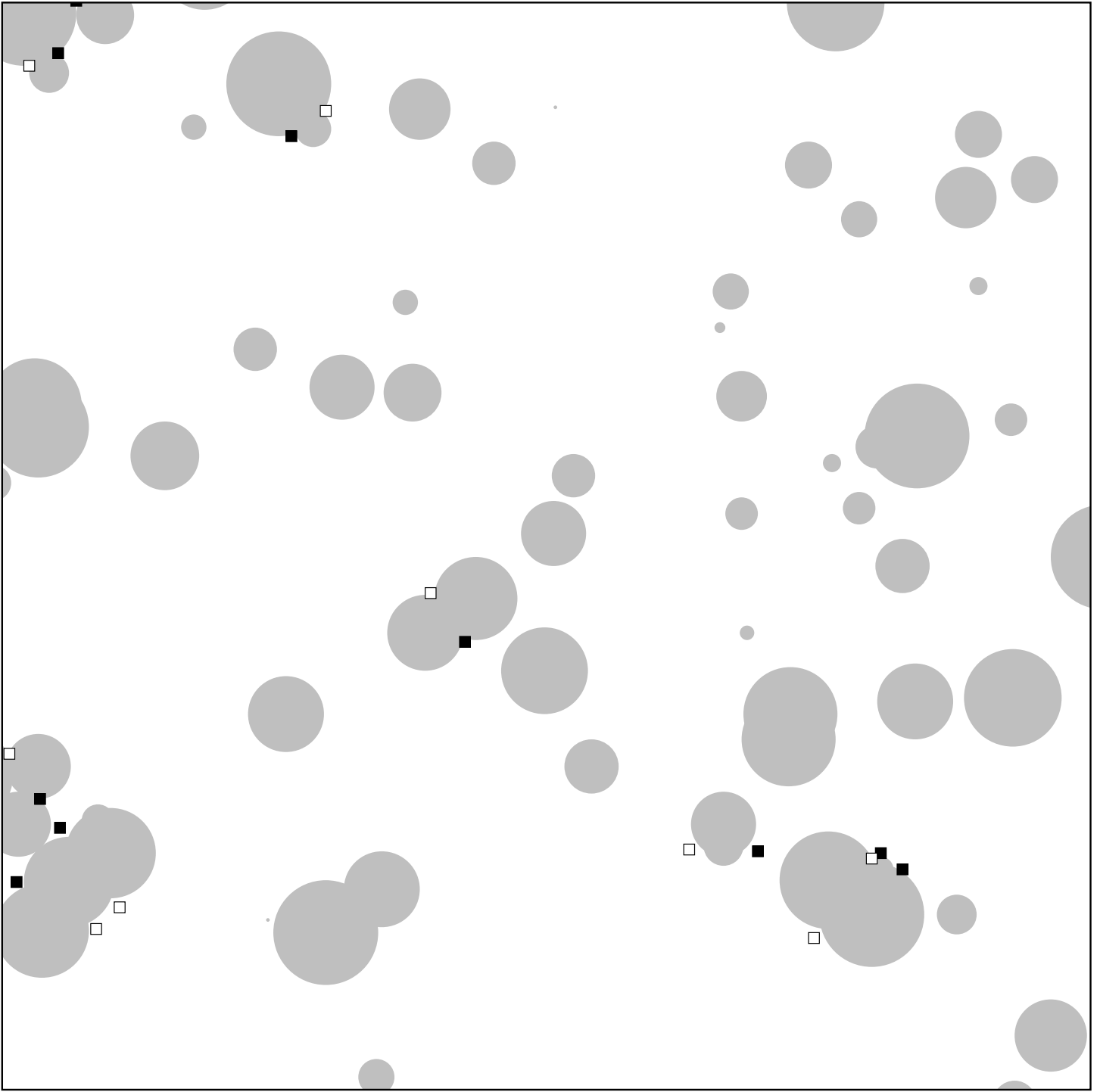
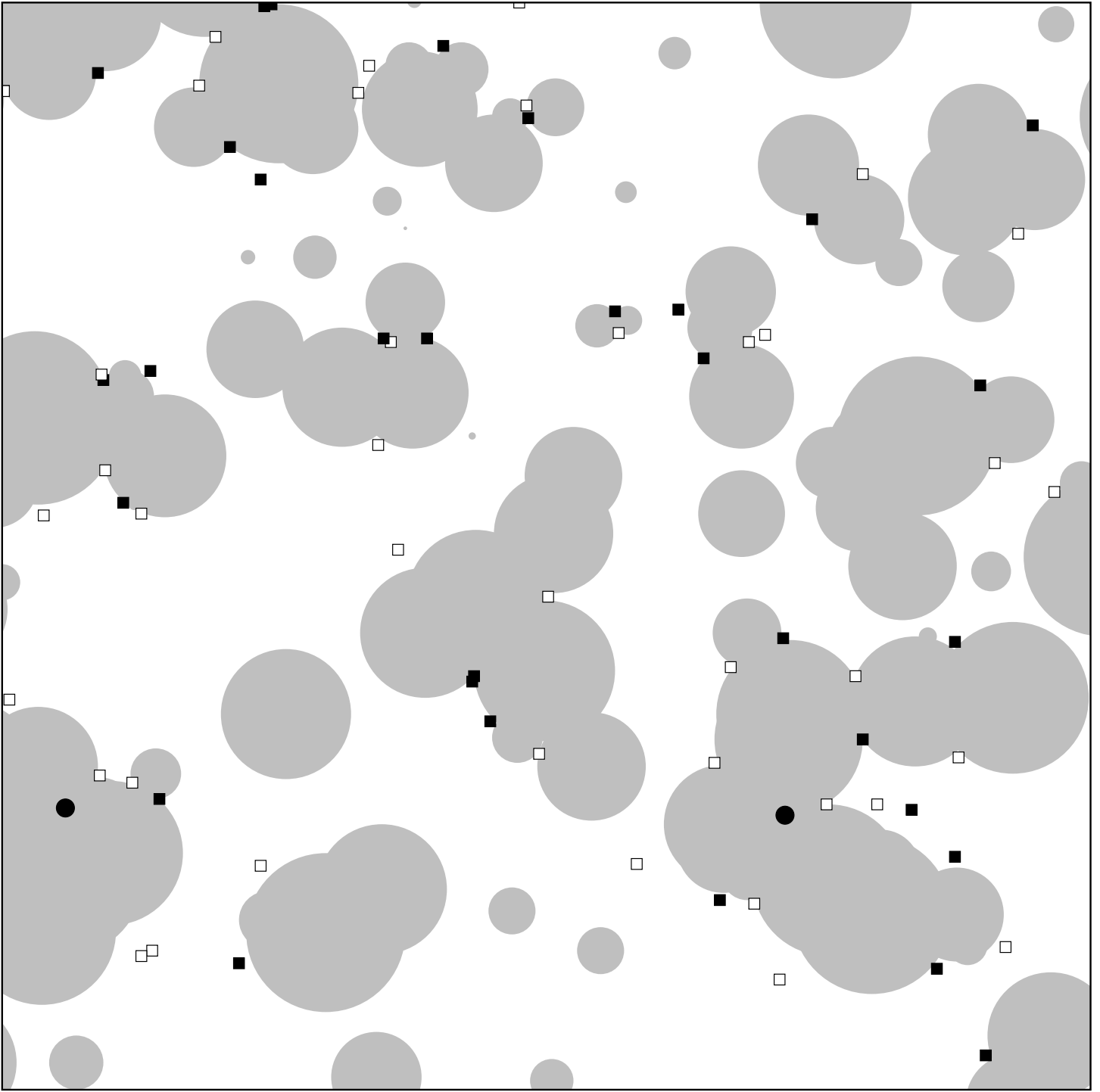
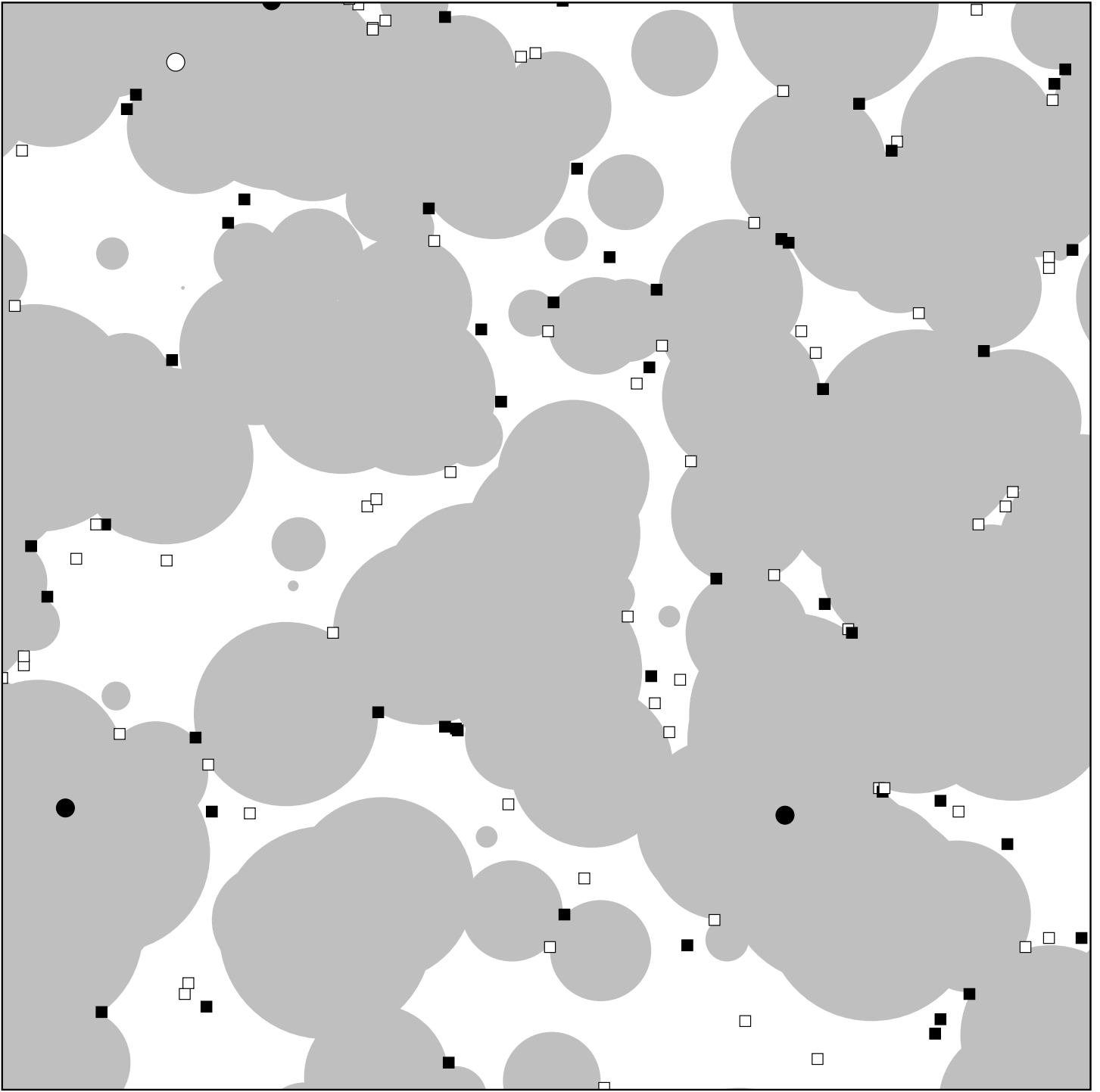
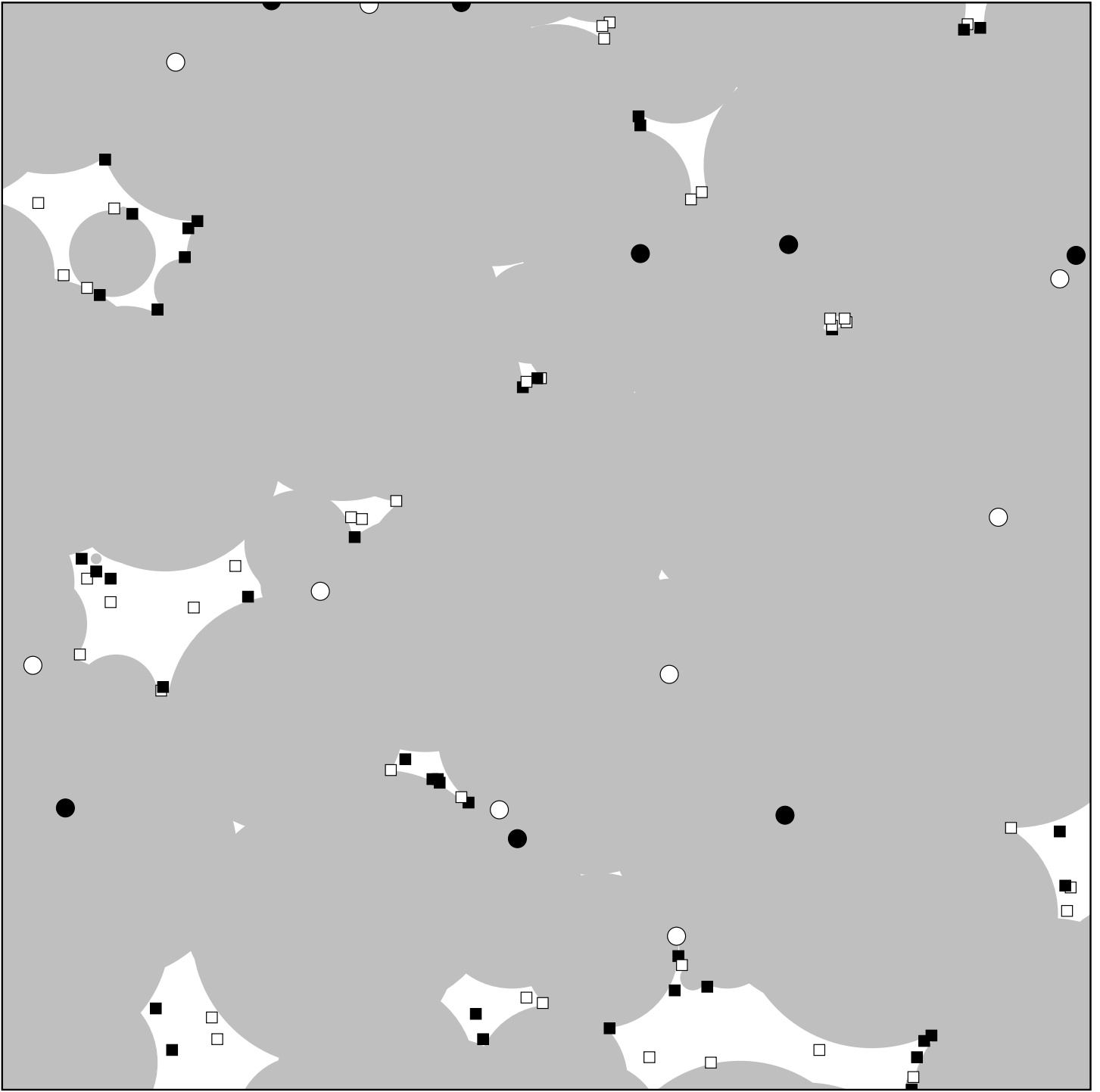
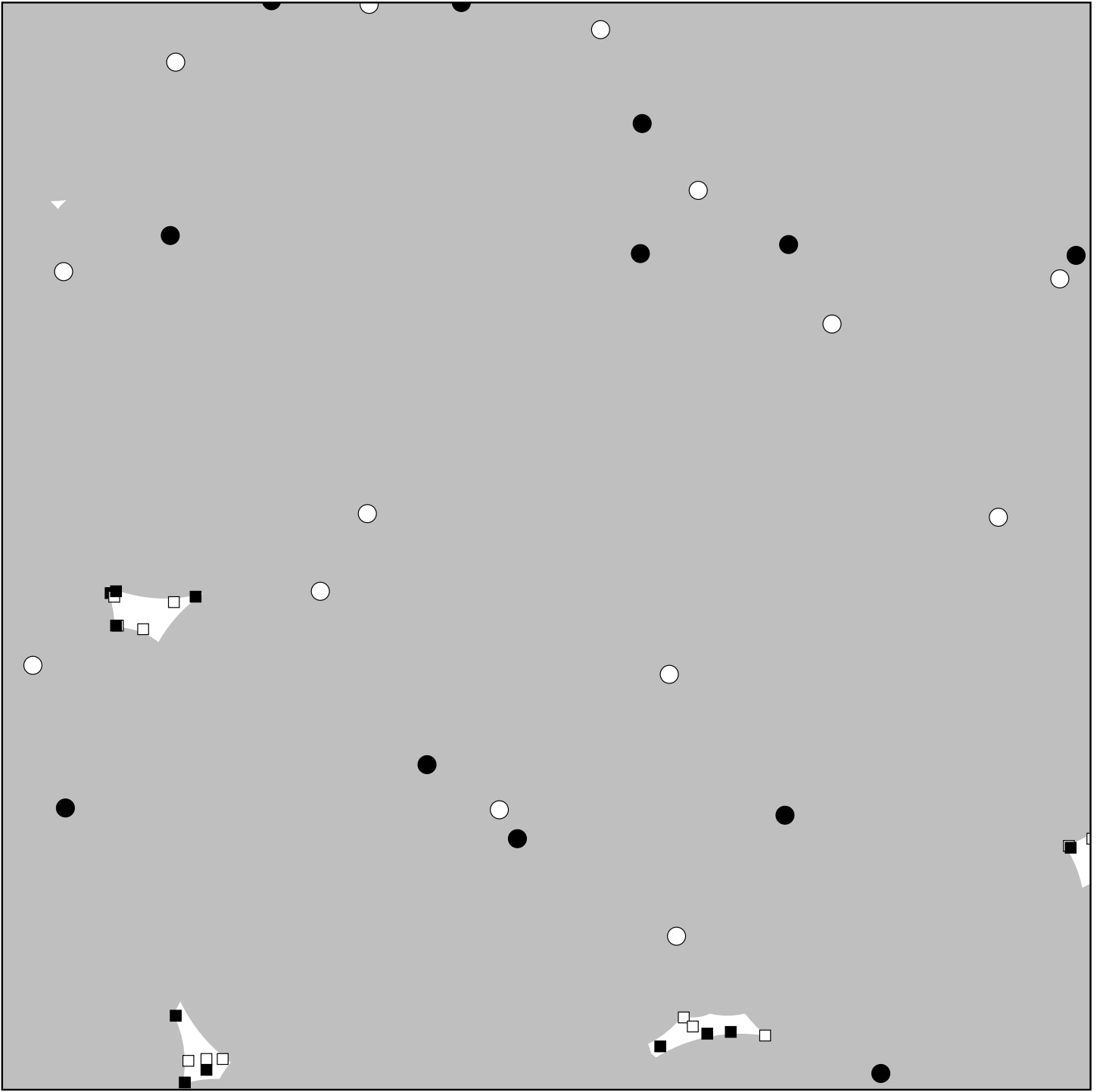
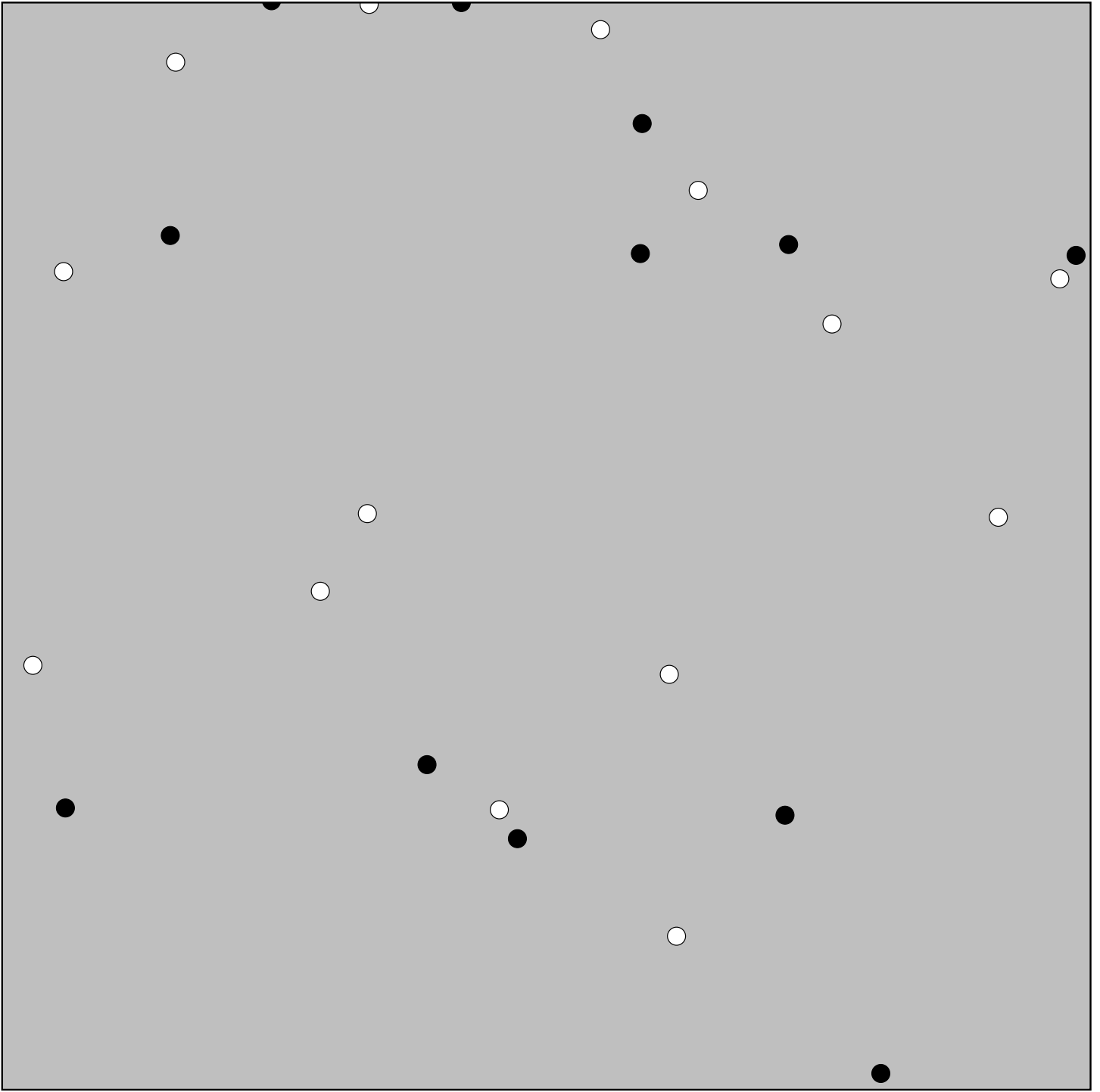
\begin{figure}

\caption{}{Snapshots of the simulation at six equally spaced times
(a)--(f) for the case $v_{\rm b} / v_{\rm f} = 0.5$.}

\end{figure}

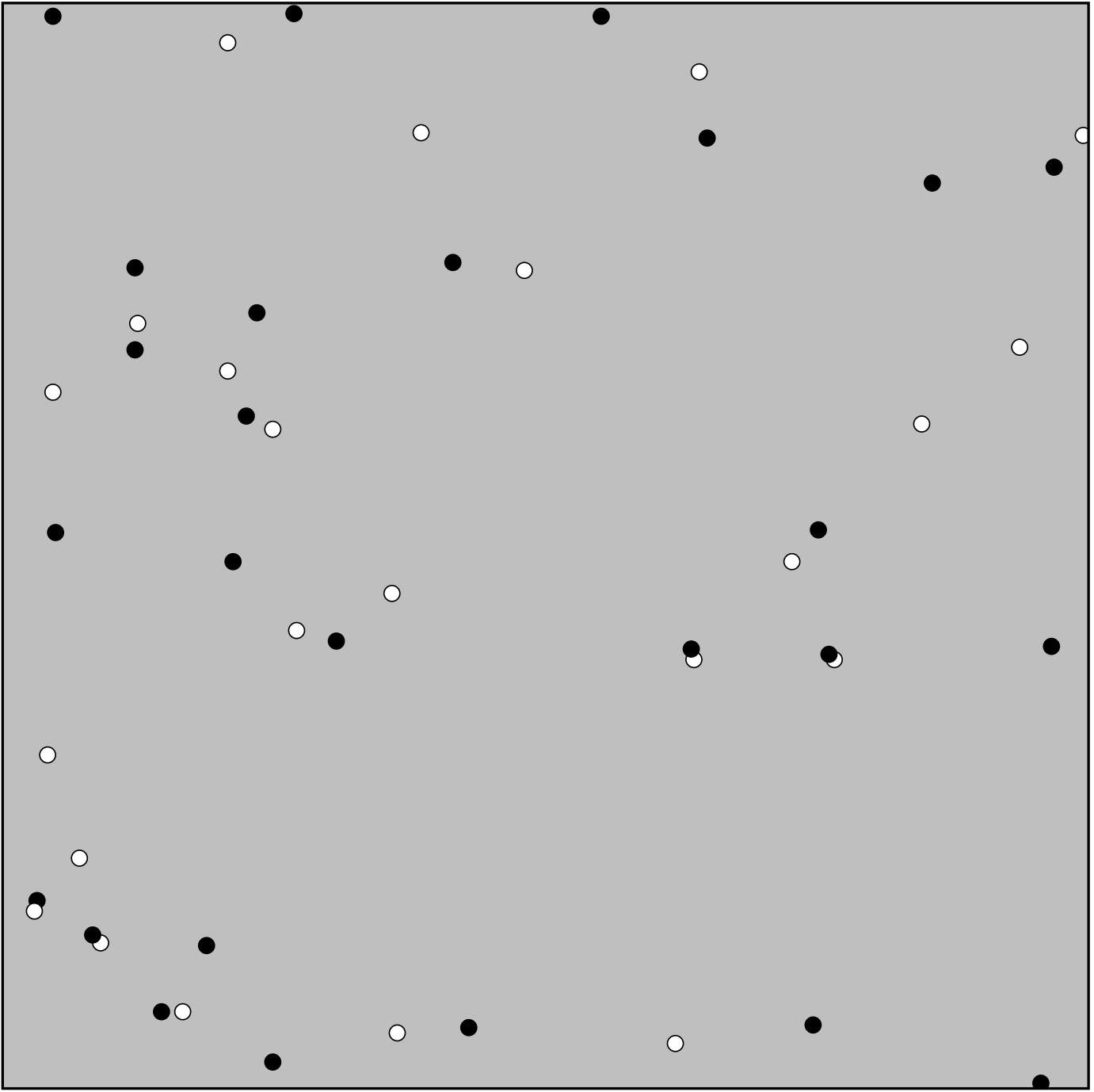
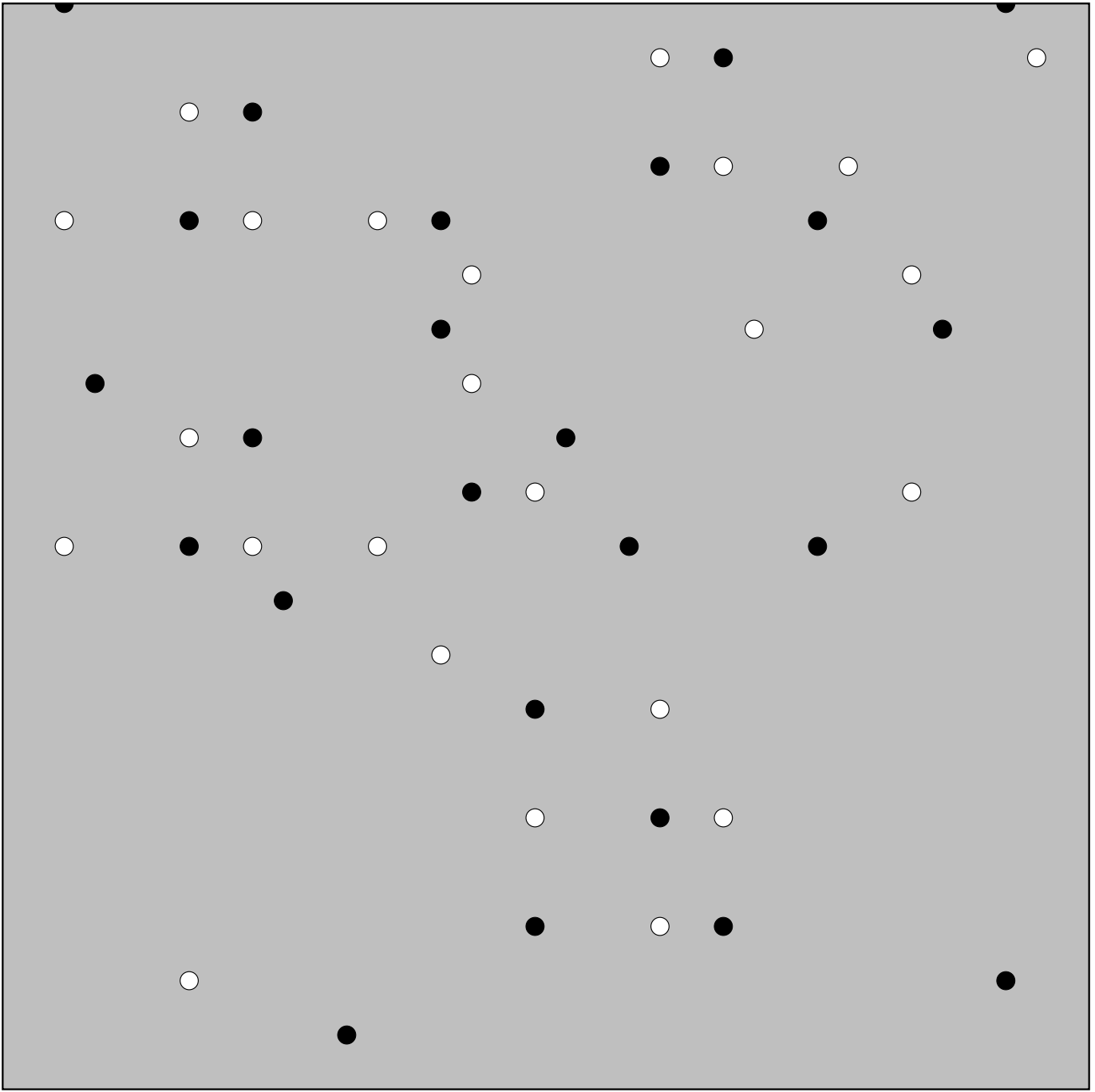
\begin{figure}

\caption{}{The final vortex distributions obtained (a) for $v_{\rm b}
/v_{\rm f} = 1$ and (b) from the standard Vachaspati-Vilenkin lattice
simulation for comparison.}

\end{figure}

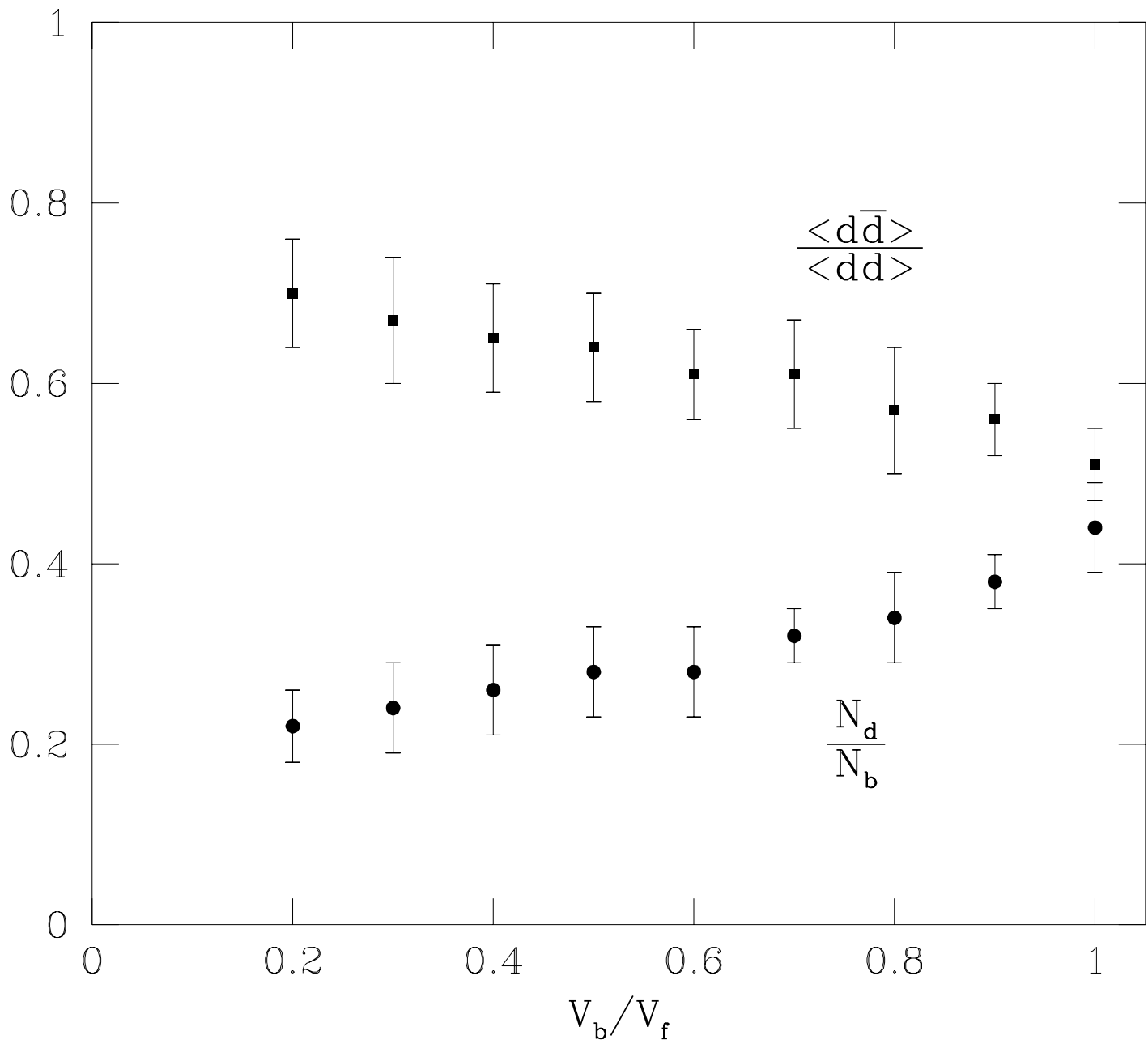
\begin{figure}

\caption{}{The mean number of defects per bubble (circles) and the
ratio between the mean nearest-neighbour defect--anti-defect and
defect--defect distances (squares) plotted against the velocity ratio
$v_b /v_f$. The error bars indicate standard deviations over 100
runs.}

\end{figure}

\end{document}